\documentclass[traditabstract]{aa}

\usepackage{amssymb}
\usepackage{amsmath}
\usepackage{graphicx}
\usepackage{natbib}
\usepackage{txfonts}
\usepackage[colorlinks=true,citecolor=blue,linkcolor=blue]{hyperref}
\usepackage{microtype}
\usepackage{color}

\bibpunct{(}{)}{;}{a}{}{,} 

\begin{document}

\title{SGR 0418+5729, Swift J1822.3-1606, and 1E 2259+586 as massive fast rotating highly magnetized white dwarfs}
\titlerunning{SGR 0418+5729, Swift J1822.3-1606, and 1E 2259 as white dwarfs}
\authorrunning{K.~Boshkayev, L. Izzo, J. A. Rueda and R. Ruffini}

\author{K.~Boshkayev, L.~Izzo, Jorge A.~Rueda, and R.~Ruffini}
\institute{Dipartimento di Fisica and ICRA, Sapienza Universit\`a di Roma, P.le Aldo Moro 5, I–-00185 
Rome, Italy\\ICRANet, P.zza della Repubblica 10, I--65122 Pescara, Italy}

\offprints{\email{kuantay@icra.it, luca.izzo@icra.it, jorge.rueda@icra.it, ruffini@icra.it}}

\date{}

\abstract{Following Malheiro et al. (2012) we describe the so-called \emph{low magnetic field magnetars}, SGR 0418+5729, Swift J1822.3--1606, as well as the AXP prototype 1E 2259+586 as massive fast rotating highly magnetized white dwarfs. We give bounds for the mass, radius, moment of inertia, and magnetic field for these sources by requesting the stability of realistic general relativistic uniformly rotating configurations. Based on these parameters, we improve the theoretical prediction of the lower limit of the spindown rate of SGR 0418+5729; for a white dwarf close to its maximum stable we obtain the very stringent interval for the spindown rate of $4.1\times 10^{-16}< \dot{P} < 6\times 10^{-15}$, where the upper value is the known observational limit. A lower limit has been also set for Swift J1822.3-1606 for which a fully observationally accepted spin-down rate is still lacking. The white dwarf model provides for this source $\dot{P}\geq 2.13\times 10^{-15}$, if the star is close to its maximum stable mass. We also present the theoretical expectation of the infrared, optical and ultraviolet emission of these objects and show their consistency with the current available observational data. We give in addition the frequencies at which absorption features could be present in the spectrum of these sources as the result of the scattering of photons with the quantized electrons by the surface magnetic field.
}


\maketitle

\section{Introduction}\label{sec:1}

Soft Gamma Ray Repeaters (SGRs) and Anomalous X-ray Pulsars (AXPs) are a class of compact objects that show interesting observational properties \citep[see e.g.][]{mereghetti08}: rotational periods in the range $P\sim (2$--$12)$ s, spindown rates $\dot{P} \sim (10^{-13}$--$10^{-10})$, strong outburst of energies $\sim (10^{41}$--$10^{43})$ erg, and in the case of SGRs, giant flares of even large energies $\sim (10^{44}$--$10^{47})$ erg.

The most popular model for the description of SGRs and AXPs, the magnetar model, based on a neutron star of fiducial parameters $M=1.4 M_\odot$, $R=10$ km and a moment of inertia $I = 10^{45}$ g cm$^2$, needs a neutron star magnetic field larger than the critical field for vacuum polarization $B_c=m^2_e c^3/(e \hbar)=4.4\times 10^{13}$ G in order to explain the observed X-ray luminosity in terms of the release of magnetic energy \citep[see][for details]{duncan92,thompson95}. There exist in the literature other models based still on neutron stars but of ordinary fields $B\sim 10^{12}$ G: these models involve either the generation of drift waves in the magnetosphere or the accretion of fallback material via a circumstellar disk \citep[see][respectively, and references therein]{2010ARep...54..925M,2013ApJ...764...49T}.

Turning to the experimental point of view, the observation of SGR 0418+5729 with a rotational period of $P=9.08$ s, an upper limit of the first time derivative of the rotational period $\dot{P} < 6.0 \times 10^{-15}$ \citep{rea10}, and an X-ray luminosity of $L_X = 6.2\times 10^{31}$ erg s$^{-1}$ can be considered as the Rosetta Stone for alternative models of SGRs and AXPs. The inferred upper limit of the surface magnetic field of SGR 0418+5729 $B<7.5\times 10^{12}$ G describing it as a neutron star \citep[see][for details]{rea10}, is well below the critical field, which has challenged the power mechanism based on magnetic field decay in the magnetar scenario.

Alternatively, it has been recently pointed out how the pioneering works of \cite{1988ApJ...333..777M} and \cite{paczynski90} on the description of 1E 2259+586 as a white dwarf can be indeed extended to all SGRs and AXPs. Such white dwarfs were assumed to have fiducial parameters $M = 1.4 M_\odot$, $R = 10^3$ km, $I = 10^{49}$ g cm$^2$, and magnetic fields $B \gtrsim 10^7$ G \citep[see][for details]{2012PASJ...64...56M} inferred from the observed rotation periods and spindown rates.

It is remarkable that white dwarfs with large magnetic fields from $10^7$ G all the way up to $10^{9}$ G have been indeed observed; see e.g.~\cite{2009A&A...506.1341K}, \cite{2010yCat..35061341K}, \cite{2010AIPC.1273...19K}, and more recently \cite{2012arXiv1211.5709K}. It is worth to mention also the fact that most of the observed magnetized white dwarfs are massive; see e.g.~REJ 0317-853 with $M \sim 1.35 M_\odot$ and $B\sim (1.7$--$6.6)\times 10^8$ G \citep[see e.g.][]{1995MNRAS.277..971B,2010A&A...524A..36K}; PG 1658+441 with $M \sim 1.31 M_\odot$ and $B\sim 2.3\times 10^6$ G \citep[see e.g.][]{1983ApJ...264..262L,1992ApJ...394..603S}; and PG 1031+234 with the highest magnetic field $\sim 10^9$ G \citep[see e.g.][]{1986ApJ...309..218S,2009A&A...506.1341K}. 

The energetics of SGRs and AXPs including their steady emission, glitches, and their subsequent outburst activities have been shown to be powered by the rotational energy of the white dwarf \citep{2012PASJ...64...56M}. The occurrence of a glitch, the associated sudden shortening of the period, as well as the corresponding gain of rotational energy, can be explained by the release of gravitational energy associated with a sudden contraction and decrease of the moment of inertia of the uniformly rotating white dwarf, consistent with the conservation of their angular momentum. 

By describing SGR 0418+5729 as a white dwarf, \cite{2012PASJ...64...56M} calculated an upper limit for the magnetic field $B < 7.5 \times 10^8$ G and show that the X-ray luminosity observed from SGR 0418+5729 can be well explained as originating from the loss of rotational energy of the white dwarf leading to a theoretical prediction for the spindown rate
\begin{equation}\label{eq:Pdotnew}
\frac{L_X P^3}{4\pi^2 I} = 1.18\times 10^{-16} \leq \dot{P}_{\rm SGR 0418+5729} < 6.0\times 10^{-15} \, ,
\end{equation}
where the lower limit was established by assuming that the observed X-ray luminosity of SGR 0418+5729 coincides with the rotational energy loss of the white dwarf. As we will show below, these predictions can be still improved by considering realistic white dwarf parameters instead of fiducial values. It is important to mention at this point that, after the submission of this work, \cite{2013arXiv1303.5579R} presented the X-ray timing analysis of the long term monitoring of SGR 0418+5729 with RXTE, SWIFT, Chandra, and XMM-Newton; which allowed the determination of the spin-down rate of SGR 0418+5729, $\dot{P}=4\times 10^{-15}$. These results confirm both our prediction given by Eq.~(\ref{eq:Pdotnew}) and the more stringent limits presented in this work in Sec.~\ref{sec:prediction} and given by Eq.~(\ref{eq:PdotminSGR0418}), which being presented in advance to the observational results presented in \citep{2013arXiv1303.5579R}, are to be considered as a predictions of the white dwarf model.

The situation has become even more striking considering the X-ray timing monitoring with Swift, RXTE, Suzaku, and XMM-Newton satellites of the recently discovered SGR Swift J1822.3--1606 \citep{rea12}. The rotation period $P=8.437$ s, and the spindown rate $\dot{P}=9.1\times 10^{-14}$ have been obtained. Assuming a NS of fiducial parameters, a magnetic field $B=2.8\times 10^{13}$ G is inferred, which is again in contradiction with a magnetar explanation for this source.

We have recently computed in \citep{2013ApJ...762..117B} general relativistic configurations of uniformly rotating white dwarfs within the Hartle's formalism \citep{1967ApJ...150.1005H}. We have used the relativistic Feynman-Metropolis-Teller equation of state \citep{2011PhRvC..83d5805R} for white dwarf matter, which we have shown generalizes the traditionally used equation of state of \cite{salpeter61}. It has been there shown that rotating white dwarfs can be stable up to rotation periods close to $0.3$ s (see \cite{2013ApJ...762..117B} and Sec.~\ref{sec:3} for details). This range of stable rotation periods for white dwarfs amply covers the observed rotation rates of SGRs and AXPs $P\sim (2$--$12)$ s.

The aim of this work is to give a detailed description of the so-called \emph{low magnetic field magnetars}, SGR 0418+5729 and Swift J1822.3-1606 as massive fast rotating highly magnetized white dwarfs. In addition to these two sources, we also present a similar analysis of the AXP prototype 1E 2259+586; which is the source on which \cite{1988ApJ...333..777M} and \cite{paczynski90} proposed the idea of a description of AXPs based on white dwarfs. We thus extend the work of \cite{2012PASJ...64...56M} by using precise white dwarf parameters recently obtained by \cite{2013ApJ...762..117B} for general relativistic uniformly rotating white dwarfs. We present an analysis of the expected Optical and near-Infrared emission from these sources within the white dwarf model and confront the results with the observational data.

\section{Rotation powered white dwarfs}\label{sec:2}

The loss of rotational energy associated with the spindown of the white dwarf is given by
\begin{equation}\label{eq:Edot}
|\dot{E}_{\rm rot}| = 4 \pi^2 I \frac{\dot{P}}{P^3} = 3.95\times 10^{50} I_{49} \frac{\dot{P}}{P^3}\quad {\rm erg s}^{-1}\, ,
\end{equation}
where $I_{49}$ is the moment of inertia of the white dwarf in units of $10^{49}$ g cm$^2$. This rotational energy loss amply justifies the steady X-ray emission of all SGRs and AXPs \citep[see][for details]{2012PASJ...64...56M}.

The upper limit on the magnetic field obtained by requesting that the rotational energy loss due to the dipole field be smaller than the electromagnetic emission of the magnetic dipole, is given by \citep[see e.g.][]{ferrari69}
\begin{equation}\label{eq:Bmax}
B=\sqrt{\frac{3 c^3}{8 \pi^2} \frac{I}{\bar{R}^6} P \dot{P}}=3.2\times 10^{15} \sqrt{\frac{I_{49}}{\bar{R}^6_8}P \dot{P}}\quad {\rm G}\, ,
\end{equation}
where $\bar{R}_8$ is the mean radius of the white dwarf in units of $10^8$ cm. The mean radius is given by $\bar{R}=(2 R_{\rm eq}+R_p)/3$ \citep[see e.g.][]{1968ApJ...153..807H} with $R_{\rm eq}$ and $R_p$ the equatorial and polar radius of the star.

It is clear that the specific values of the rotational energy loss and the magnetic field depend on observed parameters, such as $P$ and $\dot{P}$, as well as on model parameters, such as the mass, moment of inertia, and mean radius of the rotating white dwarf. It is worth mentioning that Eq.~(\ref{eq:Bmax}) gives information only on the dipole component of the magnetic field while there is the possibility that close to the star surface contributions from higher multipoles could be also important. As shown by \cite{1980NCimL..27..381Q}, the presence of higher electromagnetic multipoles increases the pulsar braking index to values larger than the traditional value $n=3$ of the magneto-dipole radiation.

\section{Structure and stability of rotating white dwarfs}\label{sec:3}

\begin{figure}[h]
\centering
\includegraphics[width=\columnwidth,clip]{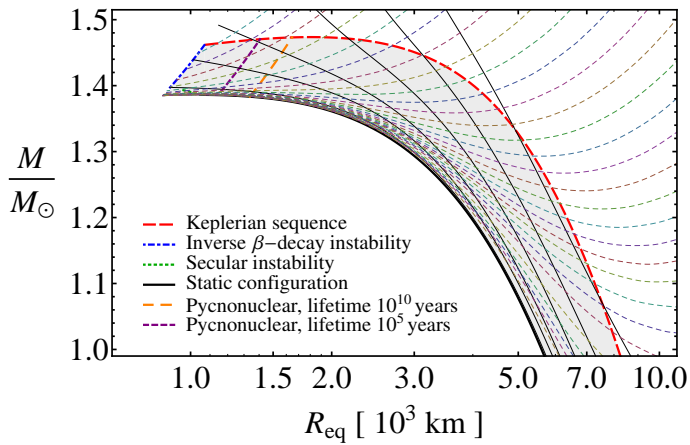}
\caption{Mass versus equatorial radius of rotating $^{12}$C white dwarfs \citep{2013ApJ...762..117B}. The solid black curves correspond to $J$=constant sequences, where the static case $J=0$ the thickest one. The color thin-dashed curves correspond to $P$=constant sequences. The Keplerian sequence is the red thick dashed curve, the blue thick dotted-dashed curve is the inverse $\beta$ instability boundary, and the green thick dotted curve is the axisymmetric instability line. The orange and purple dashed boundaries correspond to the pycnonuclear C+C fusion densities with reaction mean times $\tau_{pyc}=10$ Gyr and 0.1 Myr, respectively. The gray-shaded region is the stability region of rotating white dwarfs.}\label{fig:consOmJ}
\end{figure}

The rotational stability of fast rotating white dwarfs was implicitly assumed by \cite{2012PASJ...64...56M}. The crucial question of whether rotating white dwarfs can or not attain rotation periods as short as the ones observed in SGRs and AXPs has been recently addressed by \cite{2013ApJ...762..117B}. The properties of uniformly rotating white dwarfs were computed within the framework of general relativity through the Hartle's formalism \citep{1967ApJ...150.1005H}. The equation of state for cold white dwarf matter is based on the relativistic Feynman-Metropolis-Teller treatment \citep{2011PhRvC..83d5805R}, which generalizes the equation of state of \cite{salpeter61}. The stability of rotating white dwarfs was analyzed taking into account the mass-shedding limit, inverse $\beta$-decay and pycnonuclear instabilities, as well as the secular axisymmetric instability, with the latter determined by the turning point method of \cite{1988ApJ...325..722F}; see Fig.~\ref{fig:consOmJ} and \cite{2013ApJ...762..117B}, for details. 

The minimum rotation period $P_{\rm min}$ of white dwarfs is obtained for a configuration rotating at Keplerian angular velocity, at the critical inverse $\beta$-decay density, namely this is the configuration lying at the crossing point between the mass-shedding and inverse $\beta$-decay boundaries. The numerical values of the minimum rotation period $P_{\rm min}\approx (0.3,0.5,0.7,2.2)$ s were found for Helium, Carbon, Oxygen, and Iron white dwarfs, respectively \citep{2013ApJ...762..117B}. As a byproduct, these values show that indeed all SGRs and AXPs can be described as rotating white dwarfs because their rotation periods are in the range $2 \lesssim P \lesssim 12$ s.

The relatively long minimum period of rotating $^{56}$Fe white dwarfs, $P_{\rm min} \approx 2.2$ s, lying just at the lower edge of the observed range of rotation periods of SGRs and AXPs, reveals crucial information on the chemical composition of SGRs and AXPs, namely they are very likely made of elements lighter than Iron, such as Carbon or Oxygen.

It can be seen from Fig.~\ref{fig:consOmJ} that every $\Omega=2 \pi/P$ constant sequence intersects the stability region of general relativistic uniformly rotating white dwarfs ($M$-$R_{\rm eq}$ curves inside the shaded region of Fig.~\ref{fig:consOmJ}) in two points. These two points determine the minimum(maximum) mass $M_{\rm min,max}$ and maximum(minimum) equatorial radius $R^{\rm max,min}_{\rm eq}$, for the stability of a white dwarf rotating at the given angular velocity. Associated to the boundary values $M_{\rm min,max}$ and $R^{\rm max,min}_{\rm eq}$, we can obtain the corresponding bounds for the moment of inertia of the white dwarf, $I_{max,min}$, respectively. 

We turn now to a specific analysis of the two sources, SGR 0418+5729 and SGR SGR 1822--1606.

\section{SGR 0418+5729}\label{sec:4}

\subsection{Bounds on the white dwarf parameters}

SGR 0418+5729 has a rotational period of $P=9.08$ s, and the upper limit of the spindown rate $\dot{P} < 6.0 \times 10^{-15}$ was obtained by \cite{rea10}. The corresponding rotation angular velocity of the source is $\Omega=2\pi/P=0.69$ rad s$^{-1}$. We show in Table \ref{tab:SGR0418} bounds for the mass, equatorial radius, mean radius, and moment of inertia of SGR 0418+5729 obtained by the request of the rotational stability of the rotating white dwarf, as described in Section \ref{sec:3}, for selected chemical compositions. Hereafter we consider only general relativistic rotating Carbon white dwarfs.

\begin{table*}[tb]
\centering
{\scriptsize
\begin{tabular}{c c c c c c c c c c c}
Composition & $M_{\rm min}$ & $M_{\rm max}$ & $R^{\rm min}_{\rm eq}$ & $R^{\rm max}_{\rm eq}$ & $\bar{R}_{\rm min}$ & $\bar{R}_{\rm max}$ &
$I^{\rm min}_{48}$ & $I^{\rm max}_{50}$ & $B^{\rm upper}_{\rm min} (10^7 {\rm G})$ & $B^{\rm upper}_{\rm max} (10^8 {\rm G})$\\
\hline Helium & 1.18 & 1.41 & 1.16 & 6.88 & 1.15 & 6.24 & 3.59 & 1.48 & 1.18 & 2.90\\
Carbon & 1.15 & 1.39 & 1.05 & 6.82 & 1.05 & 6.18 & 2.86 & 1.42 & 1.19 & 3.49\\
Oxygen & 1.14 & 1.38 & 1.08 & 6.80 & 1.08 & 6.15 & 3.05 & 1.96 & 1.42 & 3.30\\
Iron & 0.92 & 1.11 & 2.21 & 6.36 & 2.21 & 5.75 & 12.9 & 1.01 & 1.25 & 0.80\\
\hline
\end{tabular}
}
\caption{Bounds for the mass $M$ (in units of $M_\odot$), equatorial $R_{\rm eq}$ and mean $\bar{R}$ radius (in units of $10^8$ cm),  moment of inertia $I$, and surface magnetic field $B$ of SGR 0418+5729. $I_{48}$ and $I_{50}$ is the moment of inertia in units of $10^{48}$ and $10^{50}$ g cm$^2$, respectively.}
\label{tab:SGR0418}
\end{table*}

\subsection{Solidification and glitches}

It has been shown by \cite{2012PASJ...64...56M} that the massive white dwarfs consistent with SGRs and AXPs possibly behave as solids since the internal temperature of the white dwarf ($\sim 10^7$ K) is very likely lower than the crystallization temperature \citep[see e.g.][]{shapirobook,usov94}
\begin{equation}\label{eq:Tcry}
T_{\rm cry} \simeq 2.3 \times 10^5 \frac{Z^2}{A^{1/3}} \left( \frac{\bar{\rho}}{10^6 {\rm g/cm}^3} \right)^{1/3} {\rm K}\, ,
\end{equation}
where $(Z,A)$ and $\bar{\rho}$ denote the chemical composition and mean density, respectively. 

This fact introduces the possibility in the white dwarf to observe sudden changes in the period of rotation, namely glitches. The expected theoretical values of the fractional change of periods of massive white dwarfs have been shown to be consistent with the values observed in many SGRs and AXPs \citep[see][for details]{2012PASJ...64...56M}.

From the bounds of $M$ and $R_{\rm eq}$ we obtain that the mean density of SGR 0418+5729 must be in the range $2.3\times 10^6 \lesssim \bar{\rho} \lesssim 5.7\times 10^8$ g cm$^3$. Correspondingly, the crystallization temperature is comprised in the range $4.8\times 10^6$ K $\lesssim T_{\rm cry} \lesssim 3.0\times 10^7$ K, where the lower and upper limits correspond to the configurations of minimum and maximum mass, respectively.

The crystallization temperature obtained here indicates that SGR 0418+5729 should behave as a rigid solid body and therefore glitches during the rotational energy loss, accompanied by radiative events, could happen. Starquakes leading to glitches in the white dwarf will occur with a recurrence time \citep[see e.g.][]{1971AnPhy..66..816B,usov94,2012PASJ...64...56M}
\begin{equation}\label{eq:tq}
\delta t_q = \frac{2 D^2}{B} \frac{|\Delta P|/P}{| \dot{E}_{\rm rot}|}\, ,
\end{equation}
where $B = 0.33\,(4 \pi/3) R^3_c e^2 Z^2 [\bar{\rho}_c/(A m_p)]^{4/3}$, $D =(3/25)\,G M^2_c/R_c$, $\dot{E}_{\rm  rot}$ is the loss of rotational energy given by Eq.~(\ref{eq:Edot}), $M_c$, $R_c$, and $\bar{\rho}_c$ are the mass, the radius and the mean density of the solid core, and $m_p$ is the proton mass.

For the minimum and maximum mass configurations and the upper limit of the spindown rate $\dot{P}<6\times 10^{-15}$, we obtain a lower limit for recurrence time of starquakes 
\begin{equation}\label{eq:EdotSGR0418}
\delta t_q  >
\begin{cases}
4.2\times 10^{9} (|\Delta P|/P)\quad {\rm yr}, & M=M_{\rm min}\\
2.0\times 10^{12} (|\Delta P|/P)\quad {\rm yr}, & M=M_{\rm max}
\end{cases}\, .
\end{equation}
For typical fractional change of periods $|\Delta P|/P=10^{-6}$, observed in SGRs and AXPs, we obtain $\delta t_q > 4\times 10^{3}$ yr and $\delta t_q >2\times 10^{6}$ yr, for $M_{\rm min}$ and $M_{\rm max}$ respectively. These very long starquake recurrent times are in agreement with the possibility that SGR 0418+5729 is an old white dwarf whose magnetospheric activity is settling down, in line with its relatively low spindown rate, magnetic field, and high efficiency parameter $L_X/\dot{E}_{rot}$, with respect to the values of other SGRs and AXPs \citep[see e.g.~Fig.~9 in][]{2012PASJ...64...56M}.

\subsection{Rotation power and magnetic field}

Introducing the values of $P$ and the upper limit $\dot{P}$ into Eq.~(\ref{eq:Edot}) we obtain an upper limit for the rotational energy loss
\begin{equation}\label{eq:EdotmaxSGR0418}
|\dot{E}_{\rm rot}| <
\begin{cases}
9.1\times 10^{32}\quad {\rm erg\,s}^{-1}, & M=M_{\rm max}\\
4.5\times 10^{34}\quad {\rm erg\,s}^{-1}, & M=M_{\rm min}
\end{cases}\, ,
\end{equation}
which for any possible mass is larger than the observed X-ray luminosity of SGR 0418+5729, $L_X = 6.2\times 10^{31}$ erg s$^{-1}$, assuming a distance of 2 kpc \citep{rea10}.

The corresponding upper limits on the surface magnetic field of SGR 0418+5729, obtained from Eq.~(\ref{eq:Bmax}) are (see also Table \ref{tab:SGR0418})
\begin{equation}\label{eq:BSGR0418}
B < B^{\rm upper}_{\rm min,max}=
\begin{cases}
1.2\times 10^{7}\quad {\rm G}, & M=M_{\rm min}\\
3.5\times 10^{8}\quad {\rm G}, & M=M_{\rm max}
\end{cases}\, .
\end{equation}

It is worth noting that the above maximum possible value of the surface magnetic field of SGR 0418+5729 obtained for the maximum possible mass of a white dwarf with rotation period $9.08$ s, $B<3.49\times 10^{8}$ G, is even more stringent and improves the previously value given by \cite{2012PASJ...64...56M}, $B<7.5\times 10^{8}$ G, based on fiducial white dwarf parameters. 

The presence of the magnetic field quantizes the electron spectrum and thus, their scattering with the photons, could generate absorption features in the spectrum at frequencies of the order of
\begin{equation}\label{eq:fcycSGR0418}
\nu_{cyc,e}=\frac{e B}{2 \pi m_e c}=
\begin{cases}
3.4\times 10^{13}\quad {\rm Hz}, & M=M_{\rm min}\\
9.8\times 10^{14}\quad {\rm Hz}, & M=M_{\rm max}
\end{cases}\, ,
\end{equation}
corresponding to wavelengths 8.9 and 0.3 $\mu$m, respectively.

\subsection{Prediction of the spindown rate}\label{sec:prediction}

Assuming that the observed X-ray luminosity of SGR 0418+5729 equals the rotational energy loss $|\dot{E}_{\rm rot}|$, we obtain the lower limit for the spindown rate
\begin{equation}\label{eq:PdotminSGR0418}
\dot{P}>\frac{L_X P^3}{4\pi^2 I}=
\begin{cases}
8.3\times 10^{-18}, & M=M_{\rm min}\\
4.1\times 10^{-16}, & M=M_{\rm max}
\end{cases}\, ,
\end{equation}
which in the case of the white dwarf with the maximum possible mass is more stringent than the value reported by \cite{2012PASJ...64...56M}, $\dot{P}=1.18\times 10^{-16}$, for a massive white dwarf of fiducial parameters.

\subsection{Optical spectrum and luminosity}

\cite{durant2011} observed SGR 0418+5729 with the two wide filters F606W and F110W of the Hubble Space Telescope, and derive the upper limits of the apparent magnitudes, $m_{F606W} > 28.6$ and $m_{F110W} > 27.4$ (Vega system), within the positional error circle derived from Chandra observations of the field of SGR 0418+5729 \citep{rea10}. The approximate distance to the source is $d = 2\pm 0.5$ kpc \citep[see][for details]{durant2011}. Assuming an interstellar extinction obtained from the $N_H$ column absorption value observed in the X-ray data, $A_V=0.7$, \cite{durant2011} obtained the corresponding luminosity upper bounds $L_{F606W} < 5\times 10^{28}$ erg s$^{-1}$ and $L_{F110W} < 6\times 10^{28}$ erg s$^{-1}$, respectively.

We use here a similar method, i.e. computing the interstellar extinction values for the V band from the $N_H$ column absorption value observed in the X-ray data, $N_H = 1.5 \times 10^{21}$ cm$^{-2}$ \citep{rea10}, and then using the empirical formula described in \citet{1995A&A...293..889P}. Then we have extrapolated the extinction to the other filters by using the method delineated in \citet{1989ApJ...345..245C}. Since the F606W and the F110W are well approximated by the V and J band, we obtained for the extinction values $A_{F606W}=0.83$ and $A_{F110W}=0.235$ respectively. The corresponding luminosity upper bounds are, consequently, $L_{F606W} < 6.82\times 10^{28}$ erg s$^{-1}$ and $L_{F110W} < 3.05\times 10^{28}$ erg s$^{-1}$. 

An estimate of the effective surface temperature can be obtained by approximating the spectral luminosity in these bands by the black body  luminosity
\begin{equation}\label{eq:Lopt}
L= 4\pi R^2 \sigma T^4\, ,
\end{equation}
where $\sigma=5.67 \times 10^{-5}$ erg cm$^{-2}$ s$^{-1}$ K $^{-4}$ is the Stefan-Boltzmann constant. For a white dwarf of fiducial radius $R=10^8$ cm, the upper limits for the surface temperature, $T<9.6\times 10^3$ K and $T<9.2\times 10^3$ K, can be obtained for the F110W and F606W filters, replacing the upper limits for $L_{F110W}$ and $L_{F606W}$ in Eq.~(\ref{eq:Lopt}). These bounds of the surface temperature of the white dwarf can be improved by using the explicit dependence on the radius of the black body surface temperature for each filter. The black body flux at a given frequency $\nu$, in the source frame, is given by
\begin{equation}\label{eq:BB}
\nu f_\nu =\pi\frac{2 h}{c^2}\frac{\nu^4}{\exp[h \nu/(k T)]-1}\, ,
\end{equation}
where $h$, $k$, and $\nu$ are the Planck constant, the Boltzmann constant, and the spectral frequency respectively. From this expression we can obtain the temperature as a function of the frequency, the observed flux, the distance $d$ and radius $R$ of the black body source
\begin{equation}\label{eq:Tbb}
T=\frac{h \nu}{k \ln\left(1 + \frac{\pi 2 h \nu^4 R^2}{c^2 d^2 F_{\nu,\rm obs}}\right)}\, ,
\end{equation}
where we have used the relation between the flux in the observed and source frames, $F_{\nu,\rm obs}=(R/d)^2\,\nu f_\nu$. 

The observed fluxes, in units of $ {\rm erg}\, ,{\rm cm}^{-2}\, ,{\rm s}^{-1}\,$, corrected for the extinction are given by
\begin{equation}\label{eq:fluxF606}
F^{F606W}_{\nu,\rm obs}=3.6 \times 10^{-20} \nu_{F606W}\times 10^{-0.4(m_{F606W}-A_{F606W})}\, ,
\end{equation}
and
\begin{equation}\label{eq:fluxF110}
F^{F110W}_{\nu,\rm obs}=1.8 \times 10^{-20}\nu_{F110W}\times  10^{-0.4(m_{F110W}-A_{F110W})}\, ,
\end{equation}
where $\nu_{F606W}=5.1\times 10^{14}$ Hz and $\nu_{F110W}=2.6\times 10^{14}$ Hz are the pivot frequencies of the F606W and F110W filters, respectively.

Introducing the upper limits of the apparent magnitudes of \cite{durant2011} with the extinction values computed in this work, Eq.~(\ref{eq:Tbb}) gives the upper bounds on the temperature

\begin{equation}\label{eq:TRconstraint}
T< 
\begin{cases}
1.3\times 10^4\,[\ln(1 + 0.44 R^2_8)]^{-1}\quad {\rm K}, & {\rm F110W}\\
2.4\times 10^4\,[\ln(1 + 6.35 R^2_8)]^{-1}\quad {\rm K}, & {\rm F606W}
\end{cases}\, ,
\end{equation}
where $R^2_8$ is the radius of the white dwarf in units of $10^8$ cm and, following \cite{durant2011}, we have approximated the band integrated flux as $\nu_c F_\nu$, with $\nu_c$ the pivot wavelength of the corresponding band filter. 

\begin{figure}[h]
\centering
\includegraphics[width=\columnwidth,clip]{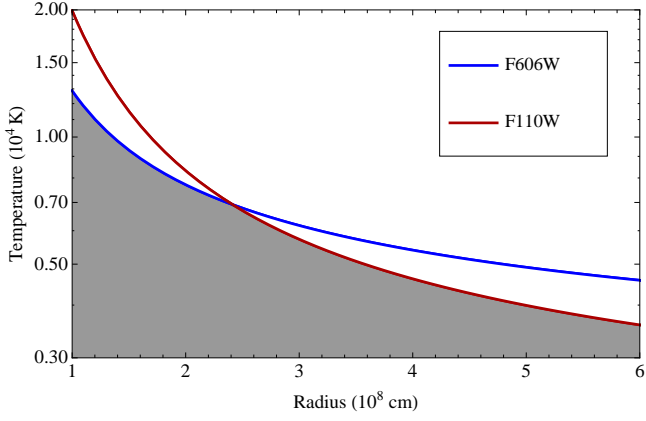}
\caption{Temperature-Radius constraint given by Eq.~(\ref{eq:TRconstraint}). The gray region corresponds to the possible values for the temperature and the radius of the white dwarf. The range of radii correspond to the one defined by the minimum and maximum mean radius of SGR 0418+5729 inferred from the white dwarf stability analysis and summarized in Table \ref{tab:SGR0418}.}\label{fig:TRconstraint}
\end{figure}

In Fig.~\ref{fig:TRconstraint}, we show the constraints on the $T$-$R$ relation obtained from Eq.~(\ref{eq:TRconstraint}). We have used the range of radii defined by the minimum and maximum radius of SGR 0418+5729 inferred from the white dwarf stability analysis and summarized in Table \ref{tab:SGR0418}.

\cite{2012PASJ...64...56M} obtained for a white dwarf of fiducial parameters the upper limits for the white dwarf surface temperature, $T<3.14\times 10^4$. We now improve these bounds on the surface temperature using realistic white dwarf parameters. From the minimum and maximum values we have obtained for the mean radius of SGR 0418+5729 (see Table \ref{tab:SGR0418}) we obtain for the F110W filter
\begin{equation}
T_{F110W} < 
\begin{cases}
4.3\times 10^3\quad {\rm K}, & M=M_{\rm min}\\
3.2\times 10^4\quad {\rm K}, & M=M_{\rm max}
\end{cases}\, ,
\end{equation}
and for the F606W filter
\begin{equation}
T_{F606W} < 
\begin{cases}
4.4\times 10^3\quad {\rm K}, & M=M_{\rm min}\\
1.2\times 10^4\quad {\rm K}, & M=M_{\rm max}
\end{cases}\, .
\end{equation}

It is clear that these constraints are in agreement with a model based on a massive fast rotating highly magnetic white dwarf for SGR 0418+5729. It is appropriate to recall in this respect some of the observed temperatures of massive isolated white dwarfs, $1.14\times 10^4\,{\rm K}\leq T \leq 5.52\times 10^4$ K as shown in the Table 1 in \citep{2005MNRAS.361.1131F}. It is also worth recalling the optical observations of 4U 0142+61 of \cite{2000Natur.408..689H} where the photometric results of the field of 4U 0142+61 at the 60-inch telescope on Palomar Mountain were found to be in agreement with a $1.3 M_\odot$ white dwarf with a surface temperature $\sim 4\times 10^5$ K \citep[see][for details]{2000Natur.408..689H}. 

\begin{figure}[h]
\centering
\includegraphics[width=\columnwidth,clip]{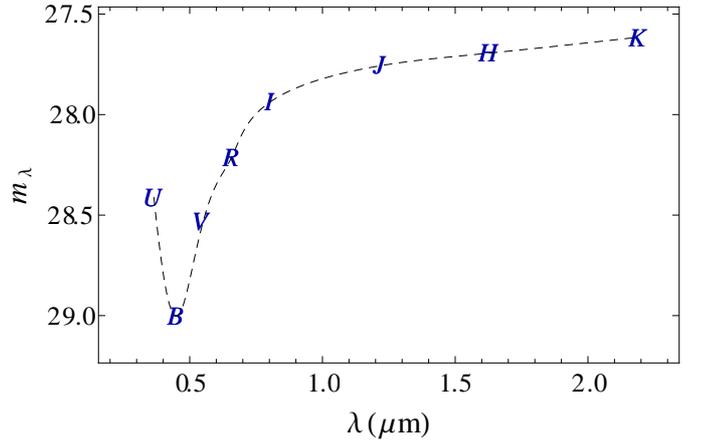}
\caption{Expected optical magnitudes of SGR 0418+5729 obtained assuming a simple blackbody for the spectral emission from a white dwarf with surface temperature $T=10^4$ K and a radius of $1.5\times 10^8$ cm, according to the constraints shown in Fig.~\ref{fig:TRconstraint}.}\label{fig:pred0418}
\end{figure}

We show in Fig.~\ref{fig:pred0418} the expected optical magnitudes of a white dwarf with surface temperature $T = 10^4$ K and radius $R = 1.5 \times 10^8$ cm, located at a distance of 2 kpc. This radius corresponds to the upper limit given by the gray region shown in Fig.~\ref{fig:TRconstraint}, for this specific value of the temperature.

\section{Swift J1822.3--1606}\label{sec:5}

\subsection{Bounds of the white dwarf parameters}

Swift J1822.3--1606 (or SGR 1822--1606) was recently discovered in July 2011 by Swift Burst Alert Telescope (BAT). A recent X-ray timing monitoring with Swift, RXTE, Suzaku, and XMM-Newton satellites found that SGR 1822-1606 rotates with a period of $P=8.44$ s and slows down at a rate $\dot{P} = 9.1 \times 10^{-14}$ \citep[see][for details]{rea12}. The corresponding rotation angular velocity of the source is $\Omega=2\pi/P=0.74$ rad s$^{-1}$. Bounds for the mass, equatorial radius, and moment of inertia of SGR 0418+5729 obtained by the request of the rotational stability of the rotating white dwarf, as described in Section \ref{sec:4}, are shown in Table \ref{tab:J1822}.

\begin{table*}[tb]
\centering
{\scriptsize
\begin{tabular}{c c c c c c c c c c c}
Composition & $M_{\rm min}$ & $M_{\rm max}$ & $R^{\rm min}_{\rm eq}$ & $R^{\rm max}_{\rm eq}$ & $\bar{R}_{\rm min}$ & $\bar{R}_{\rm max}$ &
$I^{\rm min}_{48}$ & $I^{\rm max}_{50}$ & $B_{\rm min} (10^7 {\rm G})$ & $B_{\rm max} (10^9 {\rm G})$\\
\hline Helium & 1.21 & 1.41 & 1.16 & 6.61 & 1.15 & 5.99 & 3.59 & 1.38 & 4.84 & 1.09\\
Carbon & 1.17 & 1.39 & 1.05 & 6.55 & 1.05 & 5.93 & 2.86 & 1.32 & 4.87 & 1.31\\
Oxygen & 1.16 & 1.38 & 1.08 & 6.53 & 1.08 & 5.91 & 3.05 & 1.83 & 5.80 & 1.24\\
Iron & 0.95 & 1.11 & 2.21 & 6.11 & 2.20 & 5.53 & 12.9 & 0.94 & 5.09 & 0.30\\
\hline
\end{tabular}
}
\caption{Bounds for the mass $M$ (in units of $M_\odot$), equatorial $R_{\rm eq}$ and mean $\bar{R}$ radius (in units of $10^8$ cm),  moment of inertia $I$, and surface magnetic field $B$ of Swift J1822.3--1606. $I_{48}$ and $I_{50}$ is the moment of inertia in units of $10^{48}$ and $10^{50}$ g cm$^2$, respectively.}
\label{tab:J1822}
\end{table*}

\subsection{Solidification and glitches}

The mean density of SGR 1822--1606 is in the range $2.7\times 10^6 \lesssim \bar{\rho} \lesssim 5.7\times 10^8$ g cm$^3$. The crystallization temperature for such a range following Eq.~(\ref{eq:Tcry}) is then in the range $5.0\times 10^6$ K $\lesssim T_{\rm cry} \lesssim 3.0\times 10^7$ K, which indicates that SGR 1822-1606 will likely behave as a rigid solid body.

For the minimum and maximum mass configurations and the spindown rate $\dot{P}=9.1\times 10^{-14}$, we obtain a lower limit for recurrence time of starquakes 
\begin{equation}\label{eq:EdotSGR1822}
\delta t_q  >
\begin{cases}
2.6\times 10^{8} (|\Delta P|/P)\quad {\rm yr}, & M=M_{\rm min}\\
1.1\times 10^{11} (|\Delta P|/P)\quad {\rm yr}, & M=M_{\rm max}
\end{cases}\, ,
\end{equation}
which for a typical fractional change of period $|\Delta P|/P \sim 10^{-6}$ gives $\delta t_q > 3\times 10^{2}$ yr and $\delta t_q >10^{5}$ yr, for $M_{\rm min}$ and $M_{\rm max}$ respectively. The long recurrence time for starquakes obtained in this case, confirms the similarities between SGR 1822--1606 and SGR 0418+5729 as old objects with a settling down magnetospheric activity.

\subsection{Rotation power and magnetic field}

Using the observed values of $P$ and $\dot{P}$, we obtain from Eq.~(\ref{eq:Edot}) a rotational energy loss
\begin{equation}\label{eq:EdotmaxSGR1822}
|\dot{E}_{\rm rot}| \approx
\begin{cases}
1.7\times 10^{34}\quad {\rm erg\,s}^{-1}, & M=M_{\rm max}\\
7.9\times 10^{35}\quad {\rm erg\,s}^{-1}, & M=M_{\rm min}
\end{cases}\, ,
\end{equation}
which amply justifies the observed X-ray luminosity of SGR 1822--1606, $L_X = 4\times 10^{32}$ erg s$^{-1}$, obtained assuming a distance of 5 kpc \citep[see][for details]{rea12}.

The surface magnetic field of SGR 1822.3--1606, as given by Eq.~(\ref{eq:Bmax}), is then between the values (see Table \ref{tab:J1822})
\begin{equation}\label{eq:BSGR1822}
B =
\begin{cases}
4.9\times 10^{7}\quad {\rm G}, & M=M_{\rm min}\\
1.3\times 10^{9}\quad {\rm G}, & M=M_{\rm max}
\end{cases}\, .
\end{equation}

Corresponding to the above magnetic fields, the electron cyclotron frequencies are 
\begin{equation}\label{eq:fcycSGR1822}
\nu_{cyc,e}=\frac{e B}{2 \pi m_e c}=
\begin{cases}
1.4\times 10^{14}\quad {\rm Hz}, & M=M_{\rm min}\\
3.6\times 10^{15}\quad {\rm Hz}, & M=M_{\rm max}
\end{cases}\, ,
\end{equation}
that correspond to wavelengths 2.2 and 0.08 $\mu$m, respectively. 

\subsection{Optical spectrum and luminosity}

\cite{rea12} observed the field of SGR 1822--1606 with the Gran Telescopio Canarias (GranTeCan) within the Swift-XRT position \citep{2011ATel.3493....1P}. Three sources ($S1$, $S2$, and $S3$) were detected with the Sloan $z$ filter with corresponding $z$-band magnitudes $m_{z,S1}=18.13\pm 0.16$, $m_{z,S2}=20.05\pm 0.04$, and $m_{z,S3}=19.94\pm 0.04$ \citep[see][for details]{rea12}. No additional objects were found to be consistent with the Swift-XRT position up to a magnitude $m_z=22.2\pm 0.2$ (3$\sigma$).

In addition, data from the UK Infrared Deep Sky Survey (UKIDSS) for the field of SGR 1822--1606 were found to be available, giving the magnitudes of the three aforementioned sources in the $J$, $H$, and $K$ bands; $m_{J,i}=(13.92,16.62,16.43)$, $m_{H,i}=(12.37,15.75,15.40)$, and $m_{K,i}=(11.62,15.20,14.88)$, where the index $i$ indicates the values for the sources $S1$, $S2$, and $S3$. In addition to $S1$, $S2$, and $S3$, no sources were detected within the consistent position up to the limiting magnitudes $m_{J}=19.3$, $m_{H}=18.3$, and $m_{K}=17.3$ (5$\sigma$).

\begin{figure}[h]
\centering
\includegraphics[width=\columnwidth,clip]{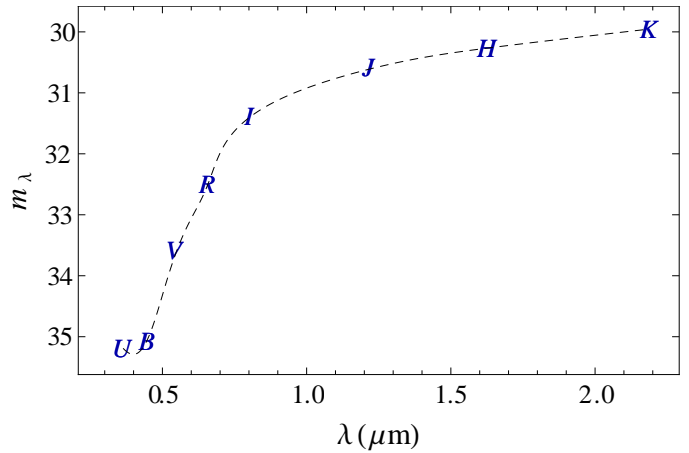}
\caption{Expected optical magnitudes of SGR 1822--1606 assuming a blackbody spectral emission from a white dwarf with surface temperature $T=10^4$ K and a radius of $1.5\times 10^8$ cm.}\label{fig:pred1822}
\end{figure}

We repeat the same analysis for SGR 0418+5729 to the case of SGR 1822--1606. We consider only the upper limits, since the three sources reported in \citet{rea12}, S1, S2 and S3, are very luminous to be a white dwarf at the distance considered for the SGR, $d \approx 5$ kpc. From the column density value, $N_H = 7 \times 10^{21}$ cm$^{-2}$, we obtain an extinction in the $V$-band of $A_V = 3.89$. From the \citet{1989ApJ...345..245C} relation we obtain the extinction values for the four bands considered, $A_z = 1.86$, $A_J = 1.10$, $A_H = 0.74$ and $A_K = 0.44$. The extinction corrected upper limits do not put very strong constraints to the temperature and the radius of the white dwarf, due to the very large distance assumed for SGR 1822--1606. We show in Fig~\ref{fig:pred1822} the expected extinction-corrected magnitudes for a white dwarf with a temperature $T = 10^4$ K and a radius $R = 1.5 \times 10^8$ cm at a distance of 5 kpc. We obtain a very deep value for the $K$-band of $\approx$ 30. We conclude that, if SGR 1822--1606 is at the distance of 5 kpc assumed by \cite{rea12}, it will be hard to detect the white dwarf. On the contrary, a possible detection would lead to a more precise determination of the distance. 

\section{1E 2259+586}\label{sec:6}

\subsection{Bounds of the white dwarf parameters}

In addition to be considered as the AXP prototype, 1E 2259+586 is the source on which \cite{1988ApJ...333..777M} and \cite{paczynski90} based their pioneering idea of describing AXPs as massive fast rotating and highly magnetized white dwarfs. This source is pulsating in X-rays with a period of P = 6.98 s \citep{fahlman81}, its spindown rate is $\dot{P} = 4.8\times 10^{-13}$ \citep{davies90} and emits X-rays with a luminosity of $L_X = 1.8\times 10^{34}$ erg s$^{-1}$ \citep{gregory80,hughes81,1988ApJ...333..777M}. The corresponding rotation angular velocity of the source is $\Omega=2\pi/P=0.90$ rad s$^{-1}$. The obtained bounds for the mass, equatorial radius, and moment of inertia of 1E 2259+586 are shown in Table \ref{tab:1E2259}.

\begin{table*}[tb]
\centering
{\scriptsize
\begin{tabular}{c c c c c c c c c c c}
Composition & $M_{\rm min}$ & $M_{\rm max}$ & $R^{\rm min}_{\rm eq}$ & $R^{\rm max}_{\rm eq}$ & $\bar{R}_{\rm min}$ & $\bar{R}_{\rm max}$ &
$I^{\rm min}_{48}$ & $I^{\rm max}_{50}$ & $B_{\rm min} (10^8 {\rm G})$ & $B_{\rm max} (10^9 {\rm G})$\\
\hline Helium & 1.28 & 1.41 & 1.15 & 5.94 & 1.15 & 5.39 & 3.56 & 1.12 & 1.26 & 2.30\\
Carbon & 1.24 & 1.39 & 1.04 & 5.88 & 1.04 & 5.34 & 2.84 & 1.08 & 1.27 & 2.76\\
Oxygen & 1.23 & 1.38 & 1.08 & 5.86 & 1.08 & 5.32 & 3.05 & 1.52 & 1.52 & 2.60\\
Iron & 1.00 & 1.11 & 2.23 & 5.49 & 2.21 & 4.98 & 13.1 & 0.78 & 1.33 & 0.62\\
\hline
\end{tabular}
}
\caption{Bounds for the mass $M$ (in units of $M_\odot$), equatorial $R_{\rm eq}$ and mean $\bar{R}$ radius (in units of $10^8$ cm),  moment of inertia $I$, and surface magnetic field $B$ of 1E2259+586. $I_{48}$ and $I_{50}$ is the moment of inertia in units of $10^{48}$ and $10^{50}$ g cm$^2$, respectively.}
\label{tab:1E2259}
\end{table*}

\subsection{Solidification and glitches}

The mean density of SGR 1822--1606 is in the range $3.9\times 10^6 \lesssim \bar{\rho} \lesssim 5.9\times 10^8$ g cm$^3$. The crystallization temperature for such a range following Eq.~(\ref{eq:Tcry}) is then in the range $5.7\times 10^6$ K $\lesssim T_{\rm cry} \lesssim 3.0\times 10^7$ K, which indicates that SGR 1822-1606 will likely behave as a rigid solid body.

For the minimum and maximum mass configurations and the spindown rate $\dot{P}=4.8\times 10^{-13}$, we obtain a lower limit for recurrence time of starquakes 
\begin{equation}\label{eq:Edot1E2259}
\delta t_q  >
\begin{cases}
4.5\times 10^{7} (|\Delta P|/P)\quad {\rm yr}, & M=M_{\rm min}\\
1.2\times 10^{10} (|\Delta P|/P)\quad {\rm yr}, & M=M_{\rm max}
\end{cases}\, ,
\end{equation}
which for a typical fractional change of period $|\Delta P|/P \sim 10^{-6}$ gives $\delta t_q > 45$ yr and $\delta t_q >1.2\times 10^4$ yr, for $M_{\rm min}$ and $M_{\rm max}$ respectively. This recurrence time for starquakes is much shorter than the ones of SGR 0418+5729 and SGR 1822--1606, indicating 1E 2259+586 as a very active source in which glitches and outburst activity, as the one observed in 2002 \citep{kaspi03,woods04}, can occur with relatively high frequency. It is interesting to note that even more frequent, with recurrence times of $\lesssim 4$ yr, can be glitches of minor intensity $|\Delta P|/P \lesssim 10^{-7}$.

\subsection{Rotation power and magnetic field} 

Using the observed values of $P$ and $\dot{P}$, we obtain from Eq.~(\ref{eq:Edot}) a rotational energy loss
\begin{equation}\label{eq:Edotmax1E2259}
|\dot{E}_{\rm rot}| \approx
\begin{cases}
1.6\times 10^{35}\quad {\rm erg\,s}^{-1}, & M=M_{\rm max}\\
6.0\times 10^{36}\quad {\rm erg\,s}^{-1}, & M=M_{\rm min}
\end{cases}\, ,
\end{equation}
much larger than the observed X-ray luminosity, $L_X = 1.8\times 10^{34}$ erg s$^{-1}$, obtained assuming a distance of $3.2\pm 0.2$ kpc \citep[see][for details]{2012ApJ...746L...4K}.

The surface magnetic field of 1E 2259+586 inferred from Eq.~(\ref{eq:Bmax}) is (see Table \ref{tab:1E2259})
\begin{equation}\label{eq:B1E2259}
B =
\begin{cases}
1.3\times 10^{8}\quad {\rm G}, & M=M_{\rm min}\\
2.8\times 10^{9}\quad {\rm G}, & M=M_{\rm max}
\end{cases}\, .
\end{equation}

Corresponding to the above magnetic fields, the electron cyclotron frequencies are 
\begin{equation}\label{eq:fcyc1E2259}
\nu_{cyc,e}=
\begin{cases}
3.6\times 10^{14}\quad {\rm Hz}, & M=M_{\rm min}\\
7.8\times 10^{15}\quad {\rm Hz}, & M=M_{\rm max}
\end{cases}\, ,
\end{equation}
that correspond to wavelengths 0.8 and 0.04 $\mu$m, respectively. 

\subsection{Optical spectrum and luminosity}

Using date from the Keck telescope, \cite{2001ApJ...563L..49H} established a faint near-IR counterpart of 1E 2259+586 with a magnitude $K_s=21.7\pm 0.2$, consistent with the position given by Chandra. In addition, upper limits in the optical bands $R=26.4$, $I=25.6$, and $J=23.8$, were placed. From the column density obtained by \cite{2001ApJ...563L..45P}, $N_H=9.3\pm 0.3 \times 10^{21}$ cm$^{-2}$, and using again the empirical formula described of \cite{1995A&A...293..889P}, one obtains the absorption $A_V=N_H/(1.79\times 10^{21}\,{\rm cm}^{-2})$, which \cite{2001ApJ...563L..49H} used to obtain the extinction in the other bands $A_R=4.3$, $A_I=3.1$, $A_J=1.4$, $A_K=0.6$.

It is known that the emission in the $K$ band, the excess in the near-IR, is typically produced by the presence of a disk; see for instance \citep{2000Natur.408..689H} for the case of 4U 0142+61 and \citep{2001ApJ...563L..49H} for the present source 1E2259+586, although in the context of a fallback disk around a neutron star. We fit the spectrum of 1E 2259+586 as the sum of a black body component
\begin{equation}
F_{\rm BB} = \pi \frac{2 h}{c^2} \left( \frac{R_{\rm WD}}{d} \right)^2 \frac{\nu^3}{e^{h \nu/(k_B T)}-1}\, ,
\end{equation}
where $R_{\rm WD}$ and $T$ the radius and effective temperature of the white dwarf, and a passive flat, opaque dust disk \citep[see e.g.][]{2003ApJ...584L..91J,2009A&A...500.1193L}
\begin{equation}
F_{\rm disk}=12 \pi^{1/3} \cos i \left( \frac{R_{\rm WD}}{d} \right)^2 \left( \frac{2 k_B T}{3 h \nu} \right)^{8/3} \left( \frac{h \nu^3}{c^2} \right) \int_{x_{\rm in}}^{x_{\rm out}} \frac{x^{5/3}}{e^x-1}dx\, ,
\end{equation}
where $i$ is the inclination angle of the disk, which we assume as zero degrees, $x_{\rm in}=h \nu/(k_B T_{\rm in})$, $x_{\rm in}=h \nu/(k_B T_{\rm out})$, with $T_{\rm in}$ and $T_{\rm out}$ the temperatures of the disk at the inner and outer radii, respectively. These temperatures are related to the radii $R_{\rm in}$ and $R_{\rm out}$ of the disk through $T_{\rm in,out}=(2/3 \pi)^{1/4}(R_{\rm WD}/R_{\rm in,out})^{3/4} T$. 

\begin{figure}[h]
\centering
\includegraphics[width=\columnwidth,clip]{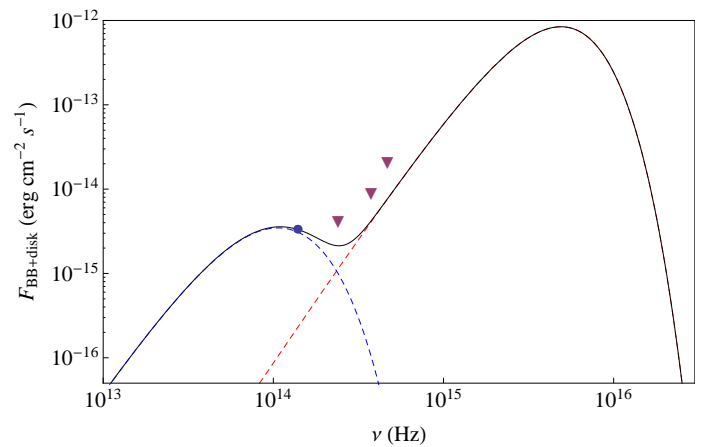}
\caption{Observed and fitted spectrum of 1E 2259+586. The filled circle is the observed flux int he $K_s$ band, and the triangles are the upper limits in the $R$, $I$, and $J$ bands. The parameters of the black body+disk spectrum are $R_{\rm WD}=3.0\times 10^8$ cm, $T=7.0\times 10^4$ K, $T_{\rm in}=2.0\times 10^3$ K, and $R_{\rm out}=R_{\odot}$. The blue-dashed curve is the contribution of the disk and the red-dashed curve is the contribution from a pure black body, the total spectrum is represented by the solid black curve.}\label{fig:fit1E2259}
\end{figure}

The total flux is then given by $F_{BB+disk}=F_{\rm BB}+F_{\rm disk}$. Since we have only one point from the observational data, the flux in the $K_s$ band, it is difficult to place constraints on the spectrum parameters. However, we can use the fact that the emission has to respect the upper limits in the $R$, $I$, and $J$ bands. We have fixed the radius of the white dwarf as $R_{\rm WD}=3\times 10^8$ cm, a value in the interval of stability of Table \ref{tab:1E2259}, and for the outer radius of the passive disk we give a typical value $R_{\rm out}=R_\odot$. We found that good fitting values of the other parameters are $T=7.0\times 10^4$ K and $T_{\rm in}=2.0\times 10^3$ K. We show in Fig.~\ref{fig:fit1E2259} the observed spectrum of 1E 2259+586 in the IR, optical, and UV bands and the composite black body + disk model spectrum. The knowledge of more data besides the $K_s$ band will lead to a definite determination of the parameter of the model and to the confirmation of the white dwarf nature of 1E 2259+586. 

It is worth to recall that the location of 1E 2259+586 appears to be associated to the supernova remnant G109.1--1.0 (CTB 109) whose age is estimated to be $t-t_0=(12$--$17)$ kyr \citep{gregory80,hughes81}. \cite{paczynski90} proposed a merger of a binary of $\sim (0.7$--$1) M_\odot$ white dwarfs that leads both to the formation of a fast rotating white dwarf and to the supernova remnant. Recent simulations of ($0.8$--$0.9 M_\odot$) white dwarf-white dwarf mergers \citep[see e.g.][]{pakmor10} point however to supernova events not very efficient energetically, below the observed explosion energy $\sim 7.4\times 10^{50}$ erg of G109.1--1.0 \citep[see e.g.][]{sasaki04}. Another interesting possibility was advanced by \cite{2012PASJ...64...56M} that this system could be originated from a tight binary system composed of a white dwarf and a late evolved companion star, approaching the process of gravitational collapse. The collapse of the companion star, either to a neutron star or to a black hole, leads to mass loss which can unbind the original binary system. If the loss of mass in the supernova explosion is $M_{\rm loss}< M/2$ with $M$ the total mass of the binary \citep[see e.g.][]{ruffini73}, the system holds bound and therefore the object will lie close to the center of the supernova remnant. Both explanations are interesting and deserve further investigation. These two scenarios may well explain the presence of a disk of material around the white dwarf, either as material expelled from the white dwarf binary merger \citep[see e.g.][]{2009A&A...500.1193L} in the scenario of \cite{paczynski90}, or as material coming from the supernova explosion and which is captured by the white dwarf in the scenario of \cite{2012PASJ...64...56M}. As we have seen, the presence of such a disk explains the emission in the near-IR observed in sources such as 4U 0142+61 \citep[see e.g.][]{2000Natur.408..689H} and in the present case of 1E 2259+586.

\section{Concluding Remarks}\label{sec:7}

We have described SGR 0418+5729, Swift J1822.3-1606, and 1E 2259+586 as massive fast rotating highly magnetized white dwarfs. The reasons for the choice of these three sources are twofold: 1) the observations of SGR 0418+5729 \citep{rea10}, $P=9.08$ and $\dot{P} < 6.0 \times 10^{-15}$, and more recently the ones of Swift J1822.3-1606 \citep{rea12}, $P=8.44$ s and $\dot{P} = 9.1 \times 10^{-14}$, challenge the description of these sources as ultramagnetized neutron stars, as required by the magnetar model; 2) 1E 2259+586 is considered the AXP prototype with very good observational data including the best example of the glitch-outburst connection \citep[see e.g.][]{woods04} and, in addition, it represents a historical object being the one analyzed by \cite{1988ApJ...333..777M} and \cite{paczynski90}, where the canonical description based on white dwarfs was proposed.

We have shown that both SGR 0418+5729 and Swift J1822.3-1606 are in full agreement with massive fast rotating highly magnetic white dwarfs. We have improved the calculation of the white dwarf parameters given by \cite{2012PASJ...64...56M} for these sources and also for 1E 2259+586. From an analysis of the rotational stability of \cite{2013ApJ...762..117B}, we have predicted the white dwarf parameters by giving bounds for the mass, radius, moment of inertia, and magnetic field of these sources; see Tables \ref{tab:SGR0418}, \ref{tab:J1822}, and \ref{tab:1E2259} for details. 

We have improved the theoretical prediction of the lower limit for the spindown rate of SGR 0418+5729, for which only the upper limit, $\dot{P} < 6.0 \times 10^{-15}$, is currently known \citep{rea10}. Based on a white dwarf of fiducial parameters, \cite{2012PASJ...64...56M} predicted for SGR 0418+5729 the lower limit $\dot{P} > 1.18\times 10^{-16}$. Our present analysis based on realistic general relativistic rotating white dwarfs allows us to improve this prediction; see Eq.~(\ref{eq:PdotminSGR0418}) in Sec.~\ref{sec:prediction} for the new numerical values. In the case of a white dwarf close to the critical mass, this new lower limit gives a very stringent constraint on the spindown rate of SGR 0418+5729, $\dot{P} = 4\times 10^{-16}< \dot{P} < 6\times 10^{-15}$, which we submit for observational verification. Indeed, after the submission of this work, \cite{2013arXiv1303.5579R} reported the confirmation of the spin-down rate of SGR 0418+5729, $\dot{P}=4\times 10^{-15}$, at 3.5 sigma confidence level. This measurement fully confirms the results of \cite{2012PASJ...64...56M}, see Eq.~(\ref{eq:Pdotnew}), as well as the more stringent constraints presented in this work, see Eq.~(\ref{eq:PdotminSGR0418}), which being presented in advance to the observations have to be considered as predictions. This fact clearly represents an observational support for the white dwarf model of SGRs/AXPs.  

In this line it is worth to mention the recent discussions on the high uncertainties and different results claimed by different authors on the value of the first period time derivative of Swift J1822.3--1606 \citep[see][and references therein for details]{2012arXiv1212.4314T}. Here we have used the value reported by \cite{rea12}. However, it would be interesting also in this case to put a theoretical lower limit with the white dwarf mode. Using $L_X = 4\times 10^{32}$ erg s$^{-1}$ at a distance of 5 kpc \citep{rea12}, we obtain a lower limit $\dot{P}\geq L_X P^3/(4\pi^2 I)\approx 2.13\times 10^{-15}$ for a $^{12}$C white dwarf close to its maximum mass; see Table \ref{tab:1E2259}. Indeed, this limit bounds from below all the observationally claimed spin-down rates for this source, known up to now.

We have given in Eqs.~(\ref{eq:fcycSGR0418}), (\ref{eq:fcycSGR1822}), and (\ref{eq:fcyc1E2259}) an additional prediction of the frequencies at which absorption features could be present in the spectrum of SGR 0418+5729, Swift J1822.3-1606, and 1E2259+586 respectively, as a result of the scattering of photons with electrons whose energy spectrum is quantized due to the magnetic field. The range we have obtained for such frequencies fall between the infrared and UV bands. In this line it is important to remark that magnetic fields in white dwarfs raging from $10^7$ G up to $10^{9}$ G are routinely observed; see e.g.~\cite{2009A&A...506.1341K}, \cite{2010yCat..35061341K}, \cite{2010AIPC.1273...19K}, and very recently \cite{2012arXiv1211.5709K} where from the Data Release 7 of the Sloan Digital Sky Survey, white dwarfs with magnetic fields in the range from around $10^6$ G to $7.3\times 10^8$ G has been found from the analysis of the Zeeman splitting of the Balmer absorption lines. Deep photometric and spectrometric observations in the range of cyclotron frequencies predicted in this work are therefore highly recommended to detect possible absorptions and line splitting features in the spectra of SGRs and AXPs.

We have also presented the optical properties of SGR 0418+5729, Swift J1822.3-1606, and 1E 2259+586 as expected from a model based on white dwarfs. In particular, the surface temperature of the white dwarf has been inferred and predictions for the emission fluxes in the UV, Optical, and IR bands have been given. We have shown that indeed the observational data are consistent with a white dwarf model for these objects. In the particular case of 1E 2259+586 the observed excess in the near-IR is explained with the presence of a disk of dust around the white dwarf. The existence of the disk can be the result of material expelled during the merger of a white dwarf binary progenitor \citep{paczynski90,ruedainprep} or as the result of material from the supernova explosion of an evolved star companion of the white dwarf in the binary scenario proposed by \cite{2012PASJ...64...56M} for the SGR/AXP-supernova connection.

It is important to discuss briefly the persistent X-ray emission of SGRs/AXPs. The time integrated X-ray spectrum is often well described by
a composite black body + power-law model with temperatures of the order of $k_{\rm B} T_{\rm BB} \sim
0.1$~keV \citep[see e.g.][]{2005A&A...433.1079G}. Such a black body component corresponds to temperatures $T_{\rm BB}\sim
10^6$ K, higher than the surface temperature of a white dwarf, as the ones predicted in this work. This clearly Spoints to an X-ray emission of magnetospheric
origin and so this black body temperature of the X-ray spectrum is not to be associated with the white dwarf effective temperature
\citep[see e.g.][]{2012PASJ...64...56M}. A possible mechanism for the X-ray quiescent emission from a magnetized white dwarf was underlined by \cite{usov93}: the reheating of the magnetosphere owing to the bombardment of the backward moving positrons created in the pair cascades formed in the interaction of the high-energy photons with the ultra-relativistic electrons. As shown by \cite{usov93} in the specific case of 1E 2259, such a reheating of the polar caps is able to produce a stable X-ray luminosity $L_X\sim 10^{35}$~erg~s$^{-1}$, in agreement with observations.

It is worth to mention that a well-known observational problem of SGRs and AXPs is the uncertainty in the estimation of the distances of the sources \citep[see e.g.][for a critical discussion on the distance of 1E2259+586]{2012ApJ...746L...4K}. These uncertainties strongly affect the estimates of the interstellar reddening $A_V$, which is crucial for the precise calculation of the source properties and therefore for a clear identification of the nature of the compact object. Deeper observations of Hubble and VLT are thus strongly recommended to establish the precise values of the luminosity in the Optical and in the near-IR bands, which will verify the white dwarf nature of SGRs and AXPs. 

We encourage future observational campaigns from space and ground to verify all the predictions presented in this work.

\begin{acknowledgements}
We thank the referee for the comments and suggestions which helped us to improve the presentation of our results.
\end{acknowledgements}


\begin{thebibliography}{55}
\expandafter\ifx\csname natexlab\endcsname\relax\def\natexlab#1{#1}\fi

\bibitem[{{Barstow} {et~al.}(1995){Barstow}, {Jordan}, {O'Donoghue},
  {Burleigh}, {Napiwotzki}, \& {Harrop-Allin}}]{1995MNRAS.277..971B}
{Barstow}, M.~A., {Jordan}, S., {O'Donoghue}, D., {et~al.} 1995, \mnras, 277,
  971

\bibitem[{{Baym} \& {Pines}(1971)}]{1971AnPhy..66..816B}
{Baym}, G. \& {Pines}, D. 1971, Annals of Physics, 66, 816

\bibitem[{{Boshkayev} {et~al.}(2013){Boshkayev}, {Rueda}, {Ruffini}, \&
  {Siutsou}}]{2013ApJ...762..117B}
{Boshkayev}, K., {Rueda}, J.~A., {Ruffini}, R., \& {Siutsou}, I. 2013, \apj,
  762, 117

\bibitem[{{Cardelli} {et~al.}(1989){Cardelli}, {Clayton}, \&
  {Mathis}}]{1989ApJ...345..245C}
{Cardelli}, J.~A., {Clayton}, G.~C., \& {Mathis}, J.~S. 1989, \apj, 345, 245

\bibitem[{{Davies} {et~al.}(1990){Davies}, {Coe}, \& {Wood}}]{davies90}
{Davies}, S.~R., {Coe}, M.~J., \& {Wood}, K.~S. 1990, \mnras, 245, 268

\bibitem[{{Duncan} \& {Thompson}(1992)}]{duncan92}
{Duncan}, R.~C. \& {Thompson}, C. 1992, \apjl, 392, L9

\bibitem[{{Durant} {et~al.}(2011){Durant}, {Kargaltsev}, \&
  {Pavlov}}]{durant2011}
{Durant}, M., {Kargaltsev}, O., \& {Pavlov}, G.~G. 2011, \apj, 742, 77

\bibitem[{{Fahlman} \& {Gregory}(1981)}]{fahlman81}
{Fahlman}, G.~G. \& {Gregory}, P.~C. 1981, Nature, 293, 202

\bibitem[{{Ferrari} \& {Ruffini}(1969)}]{ferrari69}
{Ferrari}, A. \& {Ruffini}, R. 1969, \apjl, 158, L71+

\bibitem[{{Ferrario} {et~al.}(2005){Ferrario}, {Wickramasinghe}, {Liebert}, \&
  {Williams}}]{2005MNRAS.361.1131F}
{Ferrario}, L., {Wickramasinghe}, D., {Liebert}, J., \& {Williams}, K.~A. 2005,
  \mnras, 361, 1131

\bibitem[{{Friedman} {et~al.}(1988){Friedman}, {Ipser}, \&
  {Sorkin}}]{1988ApJ...325..722F}
{Friedman}, J.~L., {Ipser}, J.~R., \& {Sorkin}, R.~D. 1988, \apj, 325, 722

\bibitem[{{G{\"o}hler} {et~al.}(2005){G{\"o}hler}, {Wilms}, \&
  {Staubert}}]{2005A&A...433.1079G}
{G{\"o}hler}, E., {Wilms}, J., \& {Staubert}, R. 2005, \aap, 433, 1079

\bibitem[{{Gregory} \& {Fahlman}(1980)}]{gregory80}
{Gregory}, P.~C. \& {Fahlman}, G.~G. 1980, Nature, 287, 805

\bibitem[{{Hartle}(1967)}]{1967ApJ...150.1005H}
{Hartle}, J.~B. 1967, \apj, 150, 1005

\bibitem[{{Hartle} \& {Thorne}(1968)}]{1968ApJ...153..807H}
{Hartle}, J.~B. \& {Thorne}, K.~S. 1968, \apj, 153, 807

\bibitem[{{Hughes} {et~al.}(1981){Hughes}, {Harten}, \& {van den
  Bergh}}]{hughes81}
{Hughes}, V.~A., {Harten}, R.~H., \& {van den Bergh}, S. 1981, \apjl, 246, L127

\bibitem[{{Hulleman} {et~al.}(2001){Hulleman}, {Tennant}, {van Kerkwijk},
  {Kulkarni}, {Kouveliotou}, \& {Patel}}]{2001ApJ...563L..49H}
{Hulleman}, F., {Tennant}, A.~F., {van Kerkwijk}, M.~H., {et~al.} 2001, \apjl,
  563, L49

\bibitem[{{Hulleman} {et~al.}(2000){Hulleman}, {van Kerkwijk}, \&
  {Kulkarni}}]{2000Natur.408..689H}
{Hulleman}, F., {van Kerkwijk}, M.~H., \& {Kulkarni}, S.~R. 2000, \nat, 408,
  689

\bibitem[{{Jura}(2003)}]{2003ApJ...584L..91J}
{Jura}, M. 2003, \apjl, 584, L91

\bibitem[{{Kaspi} {et~al.}(2003){Kaspi}, {Gavriil}, {Woods}, {Jensen},
  {Roberts}, \& {Chakrabarty}}]{kaspi03}
{Kaspi}, V.~M., {Gavriil}, F.~P., {Woods}, P.~M., {et~al.} 2003, \apjl, 588,
  L93

\bibitem[{{Kepler} {et~al.}(2010){Kepler}, {Kleinman}, {Pelisoli}, {Pe{\c
  c}anha}, {Diaz}, {Koester}, {Castanheira}, \& {Nitta}}]{2010AIPC.1273...19K}
{Kepler}, S.~O., {Kleinman}, S.~J., {Pelisoli}, I., {et~al.} 2010, in American
  Institute of Physics Conference Series, Vol. 1273, American Institute of
  Physics Conference Series, ed. {K.~Werner \& T.~Rauch}, 19--24

\bibitem[{{Kepler} {et~al.}(2012){Kepler}, {Pelisoli}, {Jordan}, {Kleinman},
  {Kulebi}, {Koester}, {Pe{\c c}anha}, {Castanheira}, {Nitta}, {da Silveira
  Costa}, {Winget}, {Kanaan}, \& {Fraga}}]{2012arXiv1211.5709K}
{Kepler}, S.~O., {Pelisoli}, I., {Jordan}, S., {et~al.} 2012, arXiv:1211.5709

\bibitem[{{Kothes} \& {Foster}(2012)}]{2012ApJ...746L...4K}
{Kothes}, R. \& {Foster}, T. 2012, \apjl, 746, L4

\bibitem[{{K{\"u}lebi} {et~al.}(2010{\natexlab{a}}){K{\"u}lebi}, {Jordan},
  {Euchner}, {Gaensicke}, \& {Hirsch}}]{2010yCat..35061341K}
{K{\"u}lebi}, B., {Jordan}, S., {Euchner}, F., {Gaensicke}, B.~T., \& {Hirsch},
  H. 2010{\natexlab{a}}, VizieR Online Data Catalog, 350, 61341

\bibitem[{{K{\"u}lebi} {et~al.}(2009){K{\"u}lebi}, {Jordan}, {Euchner},
  {G{\"a}nsicke}, \& {Hirsch}}]{2009A&A...506.1341K}
{K{\"u}lebi}, B., {Jordan}, S., {Euchner}, F., {G{\"a}nsicke}, B.~T., \&
  {Hirsch}, H. 2009, \aap, 506, 1341

\bibitem[{{K{\"u}lebi} {et~al.}(2010{\natexlab{b}}){K{\"u}lebi}, {Jordan},
  {Nelan}, {Bastian}, \& {Altmann}}]{2010A&A...524A..36K}
{K{\"u}lebi}, B., {Jordan}, S., {Nelan}, E., {Bastian}, U., \& {Altmann}, M.
  2010{\natexlab{b}}, \aap, 524, A36+

\bibitem[{{Liebert} {et~al.}(1983){Liebert}, {Schmidt}, {Green}, {Stockman}, \&
  {McGraw}}]{1983ApJ...264..262L}
{Liebert}, J., {Schmidt}, G.~D., {Green}, R.~F., {Stockman}, H.~S., \&
  {McGraw}, J.~T. 1983, \apj, 264, 262

\bibitem[{{Lor{\'e}n-Aguilar} {et~al.}(2009){Lor{\'e}n-Aguilar}, {Isern}, \&
  {Garc{\'{\i}}a-Berro}}]{2009A&A...500.1193L}
{Lor{\'e}n-Aguilar}, P., {Isern}, J., \& {Garc{\'{\i}}a-Berro}, E. 2009, \aap,
  500, 1193

\bibitem[{{Malheiro} {et~al.}(2012){Malheiro}, {Rueda}, \&
  {Ruffini}}]{2012PASJ...64...56M}
{Malheiro}, M., {Rueda}, J.~A., \& {Ruffini}, R. 2012, Publications of the
  Astronomical Society of Japan, 64, 56

\bibitem[{{Malov}(2010)}]{2010ARep...54..925M}
{Malov}, I.~F. 2010, Astronomy Reports, 54, 925

\bibitem[{{Mereghetti}(2008)}]{mereghetti08}
{Mereghetti}, S. 2008, \aapr, 15, 225

\bibitem[{{Morini} {et~al.}(1988){Morini}, {Robba}, {Smith}, \& {van der
  Klis}}]{1988ApJ...333..777M}
{Morini}, M., {Robba}, N.~R., {Smith}, A., \& {van der Klis}, M. 1988, \apj,
  333, 777

\bibitem[{{Paczynski}(1990)}]{paczynski90}
{Paczynski}, B. 1990, \apjl, 365, L9

\bibitem[{{Pagani} {et~al.}(2011){Pagani}, {Beardmore}, \&
  {Kennea}}]{2011ATel.3493....1P}
{Pagani}, C., {Beardmore}, A.~P., \& {Kennea}, J.~A. 2011, The Astronomer's
  Telegram, 3493, 1

\bibitem[{{Pakmor} {et~al.}(2010){Pakmor}, {Kromer}, {R{\"o}pke}, {Sim},
  {Ruiter}, \& {Hillebrandt}}]{pakmor10}
{Pakmor}, R., {Kromer}, M., {R{\"o}pke}, F.~K., {et~al.} 2010, Nature, 463, 61

\bibitem[{{Patel} {et~al.}(2001){Patel}, {Kouveliotou}, {Woods}, {Tennant},
  {Weisskopf}, {Finger}, {G{\"o}{\u g}{\"u}{\c s}}, {van der Klis}, \&
  {Belloni}}]{2001ApJ...563L..45P}
{Patel}, S.~K., {Kouveliotou}, C., {Woods}, P.~M., {et~al.} 2001, \apjl, 563,
  L45

\bibitem[{{Predehl} \& {Schmitt}(1995)}]{1995A&A...293..889P}
{Predehl}, P. \& {Schmitt}, J.~H.~M.~M. 1995, \aap, 293, 889

\bibitem[{{Qadir} {et~al.}(1980){Qadir}, {Ruffini}, \&
  {Violini}}]{1980NCimL..27..381Q}
{Qadir}, A., {Ruffini}, R., \& {Violini}, G. 1980, Nuovo Cimento Lettere, 27,
  381

\bibitem[{{Rea} {et~al.}(2010){Rea}, {Esposito}, {Turolla}, {Israel}, {Zane},
  {Stella}, {Mereghetti}, {Tiengo}, {G{\"o}tz}, {G{\"o}{\u g}{\"u}{\c s}}, \&
  {Kouveliotou}}]{rea10}
{Rea}, N., {Esposito}, P., {Turolla}, R., {et~al.} 2010, Science, 330, 944

\bibitem[{{Rea} {et~al.}(2012){Rea}, {Israel}, {Esposito}, {Pons},
  {Camero-Arranz}, {Mignani}, {Turolla}, {Zane}, {Burgay}, {Possenti},
  {Campana}, {Enoto}, {Gehrels}, {G{\"o}{\u g}{\"u}{\c s}}, {G{\"o}tz},
  {Kouveliotou}, {Makishima}, {Mereghetti}, {Oates}, {Palmer}, {Perna},
  {Stella}, \& {Tiengo}}]{rea12}
{Rea}, N., {Israel}, G.~L., {Esposito}, P., {et~al.} 2012, \apj, 754, 27

\bibitem[{{Rea} {et~al.}(2013){Rea}, {Israel}, {Pons}, {Turolla}, {Vigano},
  {Zane}, {Esposito}, {Perna}, {Papitto}, {Terreran}, {Tiengo}, {Salvetti},
  {Girart}, {Palau}, {Possenti}, {Burgay}, {Gogus}, {Caliandro}, {Kouveliotou},
  {Gotz}, {Mignani}, {Ratti}, \& {Stella}}]{2013arXiv1303.5579R}
{Rea}, N., {Israel}, G.~L., {Pons}, J.~A., {et~al.} 2013, \apj, in press;
  arXiv:1303.5579

\bibitem[{{Rotondo} {et~al.}(2011){Rotondo}, {Rueda}, {Ruffini}, \&
  {Xue}}]{2011PhRvC..83d5805R}
{Rotondo}, M., {Rueda}, J.~A., {Ruffini}, R., \& {Xue}, S.-S. 2011, \prc, 83,
  045805

\bibitem[{{Rueda} {et~al.}(2012){Rueda}, {Aznar-Sigu{\'a}n}, {Boshkayev},
  {Garc{\'{\i}}a-Berro}, {Izzo}, {Lor{\'e}n-Aguilar}, \&
  {Ruffini}}]{ruedainprep}
{Rueda}, J.~A., {Aznar-Sigu{\'a}n}, G., {Boshkayev}, K., {et~al.} 2012,
  submitted

\bibitem[{{Ruffini}(1973)}]{ruffini73}
{Ruffini}, R. 1973, in Black Holes (Les Astres Occlus), ed. {A.~Giannaras},
  451--546

\bibitem[{{Salpeter}(1961)}]{salpeter61}
{Salpeter}, E.~E. 1961, \apj, 134, 669

\bibitem[{{Sasaki} {et~al.}(2004){Sasaki}, {Plucinsky}, {Gaetz}, {Smith},
  {Edgar}, \& {Slane}}]{sasaki04}
{Sasaki}, M., {Plucinsky}, P.~P., {Gaetz}, T.~J., {et~al.} 2004, \apj, 617, 322

\bibitem[{{Schmidt} {et~al.}(1992){Schmidt}, {Bergeron}, {Liebert}, \&
  {Saffer}}]{1992ApJ...394..603S}
{Schmidt}, G.~D., {Bergeron}, P., {Liebert}, J., \& {Saffer}, R.~A. 1992, \apj,
  394, 603

\bibitem[{{Schmidt} {et~al.}(1986){Schmidt}, {West}, {Liebert}, {Green}, \&
  {Stockman}}]{1986ApJ...309..218S}
{Schmidt}, G.~D., {West}, S.~C., {Liebert}, J., {Green}, R.~F., \& {Stockman},
  H.~S. 1986, \apj, 309, 218

\bibitem[{{Shapiro} \& {Teukolsky}(1983)}]{shapirobook}
{Shapiro}, S.~L. \& {Teukolsky}, S.~A. 1983, {Black holes, white dwarfs, and
  neutron stars: The physics of compact objects}, ed. {Shapiro, S.~L.~\&
  Teukolsky, S.~A.}

\bibitem[{{Thompson} \& {Duncan}(1995)}]{thompson95}
{Thompson}, C. \& {Duncan}, R.~C. 1995, \mnras, 275, 255

\bibitem[{{Tong} \& {Xu}(2012)}]{2012arXiv1212.4314T}
{Tong}, H. \& {Xu}, R.~X. 2012, arXiv:1212.4314

\bibitem[{{Tr{\"u}mper} {et~al.}(2013){Tr{\"u}mper}, {Dennerl}, {Kylafis},
  {Ertan}, \& {Zezas}}]{2013ApJ...764...49T}
{Tr{\"u}mper}, J.~E., {Dennerl}, K., {Kylafis}, N.~D., {Ertan}, {\"U}., \&
  {Zezas}, A. 2013, \apj, 764, 49

\bibitem[{{Usov}(1993)}]{usov93}
{Usov}, V.~V. 1993, \apj, 410, 761

\bibitem[{{Usov}(1994)}]{usov94}
{Usov}, V.~V. 1994, \apj, 427, 984

\bibitem[{{Woods} {et~al.}(2004){Woods}, {Kaspi}, {Thompson}, {Gavriil},
  {Marshall}, {Chakrabarty}, {Flanagan}, {Heyl}, \& {Hernquist}}]{woods04}
{Woods}, P.~M., {Kaspi}, V.~M., {Thompson}, C., {et~al.} 2004, \apj, 605, 378

\end{thebibliography}
\end{document}